\documentclass{sig-alternate}

\usepackage{amssymb}
\usepackage{graphicx}
\usepackage{hanging}
\usepackage{subfigure}
\usepackage[ruled]{algorithm2e}
\usepackage{fancyvrb}
\usepackage{listings} 
\newcommand{\myfontsize}{\fontsize{7}{8.5}\selectfont}

\usepackage{url}

\makeindex

\begin{document}

\title{Efficient Incremental Breadth-Depth XML Event Mining}

\numberofauthors{1}
\author{
%
%
\alignauthor
Rashed Salem, J\'{e}r\^{o}me Darmont and Omar Boussa\"id\\
       \affaddr{Universit\'{e} de Lyon (ERIC Lyon 2)}\\
       \affaddr{5 av. P. Mend\`{e}s-France, 69676 Bron Cedex, France}\\
       \email{\{rashed.salem, jerome.darmont, omar.boussaid\}@univ-lyon2.fr}
}

\maketitle

\begin{abstract}
Many applications log a large amount of events continuously. Extracting interesting knowledge from logged events is an emerging active research area in data mining. In this context, we propose an approach for mining frequent events and association rules from logged events in XML format. This approach is composed of two-main phases: I) constructing a novel tree structure called Frequency XML-based Tree (FXT), which contains the frequency of events to be mined; II) querying the constructed FXT using XQuery to discover frequent itemsets and association rules. The FXT is constructed with a single-pass over logged data. We implement the proposed algorithm and study various performance issues. The performance study shows that the algorithm is efficient, for both constructing the FXT and discovering association rules.
\end{abstract}


\section{Introduction}
Recently, the eXtensible Markup Language (XML) has become widely used as the \textit{de facto} standard for representing, exchanging, modeling, and maintaining semi-structured data. The widespread of XML-based applications and increasing amount of XML data pose several challenges for mining XML data. Modern XML-based applications log huge amounts of events at real-time, continuously. The logged event data describe the status of each application component and can be used to trace application activities. Applications that log events in XML format range from scientific to business and financial applications. Examples of such applications include XML-based data warehousing, web personalization and web-click logs, geographic information systems, and e-commerce. Mining and analyzing logged event from such applications help for achieving self-management systems. Therefore, mining XML-formatted logged events is becoming increasingly important. It should have high attention from the database, data warehousing, data mining, and machine learning research communities.

Mining logged events is the process of extracting knowledge from continuous, rapid logged events. One of the most important data mining techniques is \textit{association rule mining}. Association rule mining discovers interesting association and/or correlation relationships among large sets of logged events, and predicts upcoming events based on occurrence of previous ones. Mining association rules from incremental XML-formatted logged events is different than mining traditional static data, due to several specific issues and challenges either related to data arrival \cite{gaber05:acm,jiang06:sigmod}, or XML-formatting nature \cite{FengDillon04,journals/dke/ZhaoCBM06}.
 
When logging events, they arrive continuously at moderate or high speed, in unbounded amount, and changing data distributions. Unlike in traditional data mining, there is not enough time to rescan the whole database whenever an update occurs. Therefore, a single-pass over events is required. Logged events need to be processed incrementally as fast as possible. Processing speed should be faster than events arrival rate. Moreover, mined data should not need to be recalculated each time requested. Unbounded amount of logged events and limited system resources, such as disk storage, memory usage, and CPU power, lead to the need for event mining algorithms that adapt themselves to available resources, otherwise accuracy result decreases. Also, while traditional data mining techniques mine frequent itemsets and discard non-frequent itemsets, this property is not valid for logged events, where the frequency of itemsets is changing over time. On the other hand, extracting knowledge from XML data is more difficult than an operational data, because of the flexible, irregular, and semi-structured nature of XML data.

To the best of our knowledge, there is no algorithm proposed in the literature to discover interesting knowledge from incremental XML-formatted logged events. Therefore, we propose in this paper an incremental algorithm for this purpose. Our algorithm is composed of two main phases: firstly, we construct a new tree structure called Frequency XML-based Tree (FXT) that stores frequencies of events to be mined. Secondly, we query frequent event-sets and association rules efficiently from the constructed FXT using XQuery. Our algorithm handles most processing logged event issues. It satisfies a single-pass over data transactions to construct the compact FXT structure. Although the FXT is processed using XML technologies and constructed in XML format, its construction time is fast enough. Association rules with different minimum supports are queried at any time without re-constructing the FXT from scratch.

The rest of this paper is organized as follows. Related work is discussed in section \ref{sect:related}. In section \ref{sect:loggedEvents}, we present our motivation and a description of logged events. Section \ref{sect:fxt} introduces the general structure of the novel Frequency XML-based Tree (FXT) and our algorithm for constructing the FXT. Mining frequent itemsets and association rules from the FXT is presented in section \ref{sect:miningAR}. Performance study of our algorithm is discussed in section \ref{sect:performance}. Finally, we conclude and highlight future trends in section \ref{sect:conclusion}. 
\section{Related Work}
\label{sect:related}

There are two main types of approaches for XML data mining in the literature. The first type of approaches applies relational data mining tools on XML data by mapping XML documents to relational data model and storing them in a relational database \cite{zhangJi06:jdm}.
The second type of approaches applies data mining techniques directly onto native-XML data \cite{braga03:ictai,wan04:dmwi,wang08:alpit}. 
We are interested with the second type of approaches, specifically mining frequent itemsets and association rules from XML data. 

\paragraph{\textbf{Mining association rules using XQuery}} Wan and Dobbie provide XQuery implementation of the well-known Apriori algorithm \cite{agrawal94:vldb}, to extract association rules from XML documents without any pre-processing or post-processing \cite{wan04:dmwi}. Their algorithm is adapted to simple and well-defined XML format. This algorithm is extended with pre-processing step in order to mine more complex and irregular XML documents \cite{wang08:alpit}. Authors actually transform complex documents into a format that can be mined by Wan algorithm using XSLT. 
Braga et al. propose XMINE \cite{braga03:ictai}, a tool to extract XML association rules from XML documents. The XMINE operator is based on XPath and XQuery to express complex mining tasks on the content and the structure of XML data. 

\paragraph{\textbf{Tree-based mining algorithms}} Han et al. propose FP-Growth for mining frequent itemsets without generating candidate itemsets \cite{han00:sigmod}. FP-Growth requires two database scans for constructing its FP-Tree. 
Cheung and Zaiane extend FP-Tree by proposing a novel data structure called CATS Tree \cite{cheung03:ideas}. As FP-Tree, CATS tree allows frequent pattern mining without generation of candidate itemsets. It allows mining with a single pass over the database as well as efficient insertion or deletion of transactions at any time. 

To the best of our knowledge, our algorithm is the first work proposed to mine frequent itemsets and association rules from incremental XML-formatted logged events, using XML technologies (e.g., XPath and XQuery). Table \ref{tab:comparison} shows differences between FXT versus tree-based techniques (i.e., FP-Growth and CATS) and XQuery-based implementation techniques (i.e., Apriori implementation). Although Apriori-implementation mines association rules from XML data using XQuery \cite{wan04:dmwi}, it is designed  to static transactions of XML data. Mining association rules with different minimum support by Apriori algorithm requires re-generating the largest itemsets from scratch. Compared to our algorithm, Apriori-implementation provides less performance particularly for large databases of transactions. Despite CATS \cite{cheung03:ideas} is not proposed for mining XML data, it is based on constructing an incremental frequency tree like our algorithm. 
Rather than CATS algorithm mines frequent patterns with a complicated algorithm named FELINE, it does not support mining association rules from CATS tree directly due to the absence of total size of transactions. 

\begin{table}[htbp]
	\centering
		\begin{tabular}{c}
		
		\includegraphics[width=\columnwidth]{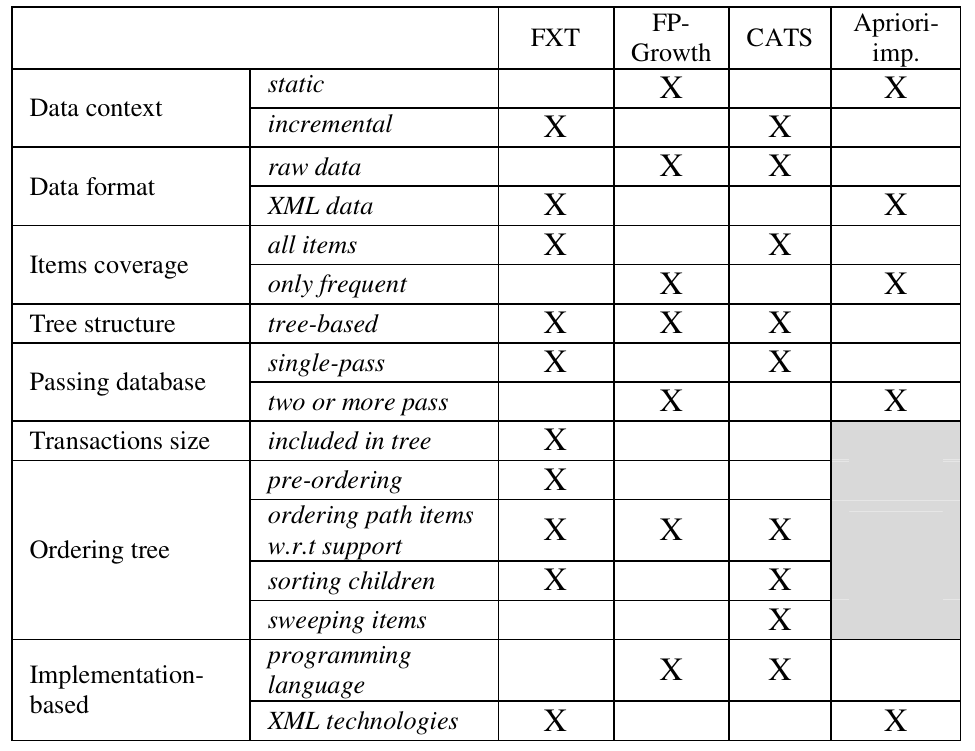}
			
		\end{tabular}
	\caption{FXT versus related works}
	\label{tab:comparison}
\end{table}

\section{Logging Events}
\label{sect:loggedEvents}

\begin{figure}[htpb]
\centering
\includegraphics[width=7.5cm]{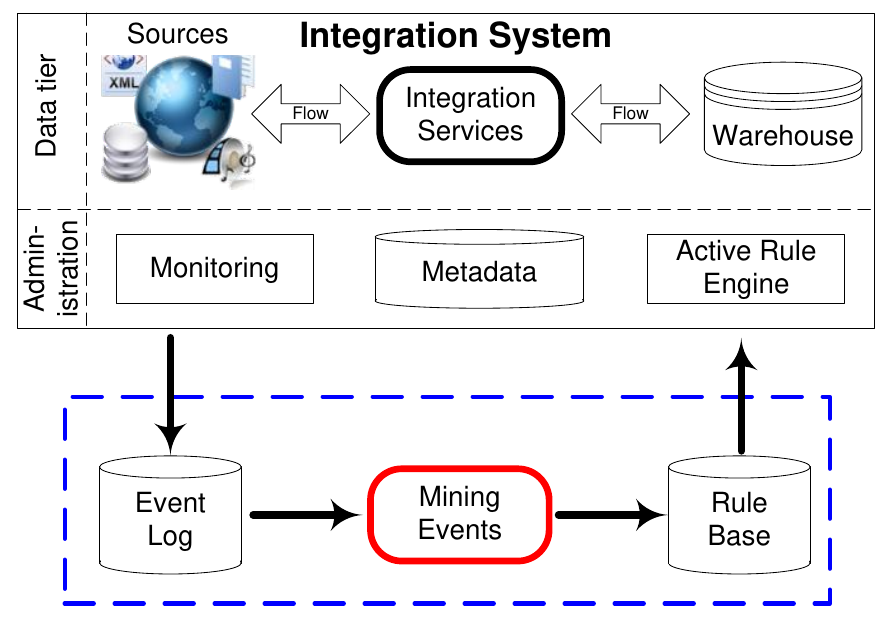}
\caption{Data integration system}
\label{fig:data_integration}
\end{figure}

There are several software platforms that log a large amount of events incrementally every day, into simple text or XML format. Logged events are essential to understand and trace the activities of such platforms. For instance, we are motivated to mine logged events from XML-based data integration platforms \cite{salem10:icwmi}. It worth to be noted that these platforms are developed, managed and maintained using XML technologies. Data integration is the process of extracting data from heterogeneous and distributed sources, transforming them into a unified format, and loading them into a repository (namely a warehouse), see Figure \ref{fig:data_integration}. Discovering interesting knowledge from logged events can be employed to self-maintain and configure the workflow behavior of these systems, how to achieve this issue is out scope of this paper.

Actually, logged events include much descriptive information about each activity (e.g., identification, occurring time, source, description, category, etc.). The more logging information, the more interesting knowledge can be discovered. In order to apply our algorithm for mining frequent events and association rules, we need to pre-process logged events and organize them into transactions as illustrated in the following sample. 

\vspace{30pt}

{
\myfontsize
\begin{Verbatim}[frame=lines, xleftmargin=5mm]
<transaction id="1" time="2011-04-10 09:16:00">
 <item>A</item> <item>B</item> <item>C</item> <item>D</item>
</transaction>
<transaction id="2" time="2011-04-10 09:16:20">
 <item>C</item> <item>E</item>
</transaction>
<transaction id="3" time="2011-04-10 09:16:40">
 <item>B</item> <item>C</item>
</transaction>
<transaction id="4" time="2011-04-10 09:17:00">
 <item>C</item> <item>D</item> <item>E</item>
</transaction>
<transaction id="5" time="2011-04-10 09:17:20">
 <item>B</item> <item>C</item> <item>D</item>
</transaction>
<transaction id="6" time="2011-04-10 09:17:40">
 <item>A</item> <item>C</item> <item>E</item>
</transaction>
		\end{Verbatim}	
}

Each transaction has identification, its occurring time, and a set of items which represent platform events. The set of events is assumed to be logged in a window-size of time (time + window). The corresponding format of logged transactions can be obtained directly from their origin platforms, or can be transformed using the XSLT language to a format that can be processed by our algorithm in a pre-processing step. The most important thing to our algorithm is to define the listing of items of each transaction, which should be sorted alphabetically for performance purposes.

\section{Frequency XML-based Tree (FXT)}
\label{sect:fxt}
\subsection{FXT Structure}

\begin{figure}[htpb]
\centering
\includegraphics[width=\columnwidth]{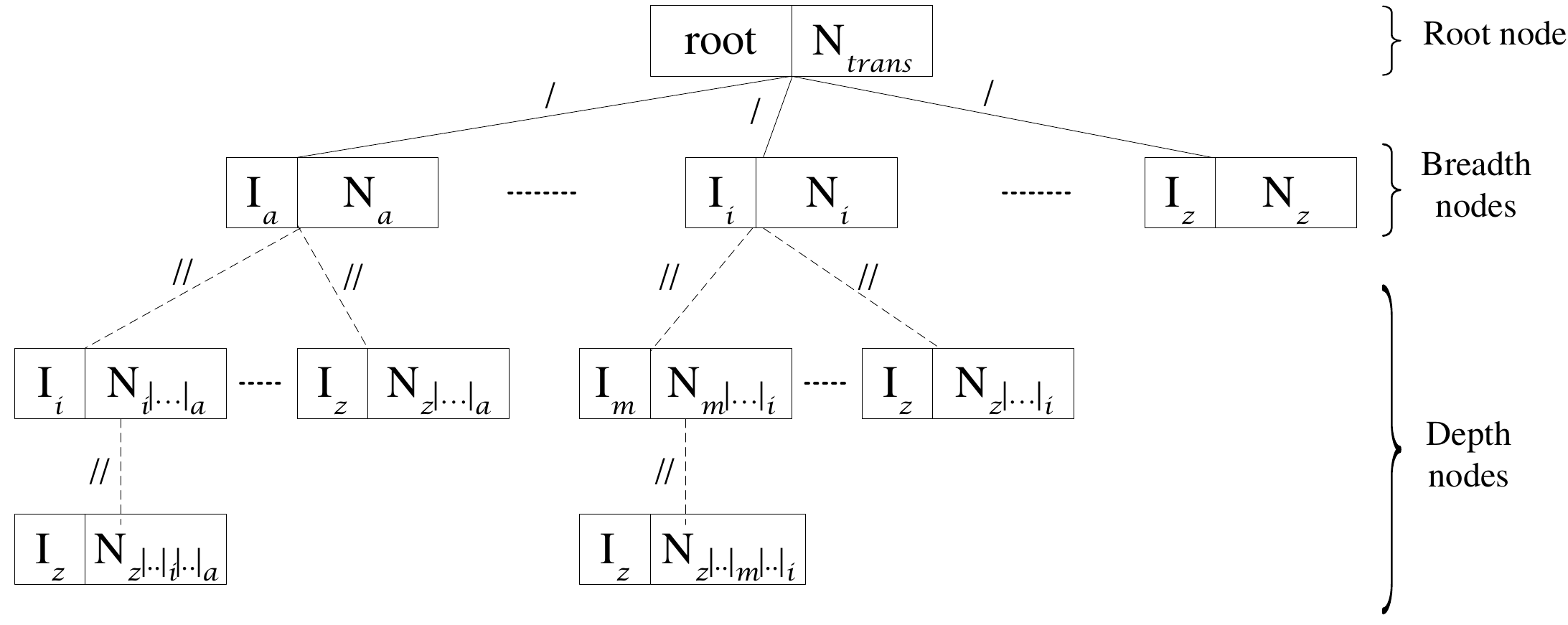}
\caption{FXT structure}
\label{fig:fxt_structure}
\end{figure}

In order to mine frequent itemsets or association rules, the frequency of events (or items) needs to be calculated. Hence, we propose a novel tree structure that contains frequency of all logged items, named Frequency XML-based Tree (FXT). The FXT nodes, except root node, consist of two entries: \textit{item name} and \textit{counter}, where \textit{item name} registers which item this node represents (e.g., $I_i$), and \textit{counter} registers the number of transactions represented by the portion of the path reaching this node (e.g., $N_i$ or $N_{m|...|i}$). As illustrated in figure \ref{fig:fxt_structure}, the FXT is composed of three main levels of nodes. Firstly, the \textit{Root node} refers to the FXT root node. It represents the total number of logged transactions ($N_{trans}$). Secondly, the \textit{Breadth nodes} refers to all root's children nodes. It represents the count of each item appeared in any logged transaction. 
Thirdly, the \textit{Depth nodes} refers to all root's grandchildren nodes. It represents a relative or conditional count of a specific item given other related items. 
The depth nodes are represented as set of paths, each path corresponds specific transactions itemsets. In figure \ref{fig:fxt_structure}, the 
dashed line annotated by double slashes ``//" means that there is zero or more in-between nodes in a specific depth path. 

It worth to be noted that the FXT can handle both sorted and unsorted items of upcoming transactions, but we observed that handling sorted items results in more compact FXT structure and eases mining frequent itemsets and association rules from the FXT. Thus, letters (\textit{a, i, m,} and \textit{z}) of items 
refer to their ordering. 
In addition, although FXT is designed to manage XML-formatted data, the same concept can be applied to raw data. Finally, there are some facts can be deduced from the FXT structure:\begin{itemize}
	\item $N_{trans} = Total(trans)$ refers to the total number of transactions; 
  \item $N_{trans} \geq N_k$, where $N_k$ can be count of any item $k$;
	\item $N_k \geq N_{v|...|k}$, where $N_{v|...|k}$ is a conditional count of $I_v$ given $I_k$ and in-between items. 
\end{itemize}

\subsection{FXT Management}

The first phase of our algorithm is to construct the FXT, by handling each logged transaction individually. 

\subsubsection{Insertion of transactions}

 Logged transactions are inserted into the FXT upon arrival. 
Our algorithm follows four steps for each logged transaction on constructing the FXT as presented by algorithm \ref{alg:main}.

\paragraph{\textbf{Step 1 }(\textit{incrementing root counter}, $N_{trans}$)} This root counter represents the total number of logged transactions, which can be used to calculate item support.

\begin{algorithm}[htpb]
\SetAlCapFnt{\small\sf}
\small
\DontPrintSemicolon
\textbf{Input:} Set of transactions (S) \\
\textbf{Output:} FXT document  \\
\Begin{
  \ForEach{$T \in S$ }{
			idx = 0  \\
			\textit{(: step1: increment root counter :)}\\
			root/@counter++   \\    			
			increment-or-create-breadth(T)\\
			increment-or-create-depth(T)\\
			update-other-paths(T, idx )
	  } 
	}
\caption{FXT construction\label{alg:main}}
\end{algorithm}

\paragraph{\textbf{Step 2 }(\textit{incrementing breadth})} For each item of the transaction, our algorithm increments the item counter if it exists as one of root children (breadth nodes), otherwise the algorithm creates the item as new root child and initializes its counter at 1. 
Any \textit{item} support can be easily calculated later via dividing item counter by $N_{trans}$, see algorithm \ref{alg:step2}.

\begin{algorithm} [htpb]
\SetAlCapFnt{\small\sf}
\small
\DontPrintSemicolon
\textbf{Procedure:} increment-or-create-breadth(T)\\
\textbf{Input:} Transaction(T)  \\
\Begin{
  \ForEach{$item \in T$ }{
		\eIf{$item \in root/*$}{
				item/@counter++	              
		}{
				\textit{(: create new item as root child, initialize its counter at 1 :)} \\
				insert root/item					\hspace{0.4in}\textit{(: item as root child :)} \\
				item/@counter=1
		  }
	  }
}
\caption{Incrementing breadth\label{alg:step2}}
\end{algorithm}

\paragraph{\textbf{Step 3 }(\textit{incrementing depth})} The algorithm increments the transaction path if it exists, otherwise creates it. While creating the path, item by item, respecting the transaction items ordering, the algorithm takes into account the previous occurrence of relative transaction items. This is reflected when initiating counter of the path items. The FXT path may not correspond only the same transaction that occurred once or several times, but also correspond many transactions that satisfy the same beginning portion of the path. This step is presented by
algorithm \ref{alg:step3}.



\begin{algorithm} [htbp]
\SetAlCapFnt{\small\sf}
\small
\DontPrintSemicolon
\textbf{Procedure:} increment-or-create-depth(T)\\
\textbf{Input:} Transaction(T)  \\
\Begin{
	path = root  \\
	preItemPaths = 0  \\
  \For{idx = 0 to length(T)-1}{
		path $\leftarrow$ path / item[idx]        \hspace{25pt} \textit{(:where item is T member:) } \\ 
		nexIdx $\leftarrow$ idx+1					\\
		nextItem $\leftarrow$ item[nexIdx]                          	        \hspace{25pt} \textit{(:next item in T:)} \\
		\textit{(: paths that have the item anywhere: )}  \\
		preItemPaths $\leftarrow$ preItemPaths // item[idx]  \\ 
		\textit{(:if inserted item is child of its previous item path:)}\\							
		\uIf{nextItem $\in$ path/*}													   
		  {	nextItem/@counter++  \\
			\textit{(: if next item is descendant of its previous items :)} \\
		 }
		\uElseIf{nextItem $\in$ root//preItemPaths//*}{			
					insert path/nextItem					\\
				 \textit{(: max preItem counter is incremented as initialization counter :)}  \\
					nextItem/@counter $\leftarrow$ Max(root//preItemPaths//nextItem/@counter)+1 \\
		 }				
		\Else{
						insert path/nextItem   \\
						nextItem/@counter=1
					}
			}
}
\caption{Incrementing depth\label{alg:step3}}
\end{algorithm}

\paragraph{\textbf{Step 4 }(\textit{updating other paths})} This step is required to ensure the correctness of counting of one given itemset across different FXT paths. For each transaction, some paths can be generated from transaction items that differ from the path built in step 3, called \textit{other paths}. The algorithm checks only other paths existing in the FXT to be updated. In case if they do not already exist, the algorithm does not create them for compactness purpose,  
see algorithm \ref{alg:step4}. 
\begin{algorithm}[htbp]
\SetAlCapFnt{\small\sf}
\small
\DontPrintSemicolon
\textbf{Procedure:} update-other-paths(T, idx ) \\
\textbf{Input:} Transaction(T), index of breadth item (idx) \\
\Begin{
	path = root  \\
  \For{idx to length(T)-1}{
		path $\leftarrow$ path / item[idx] 		\hspace{0.2in} \textit{\footnotesize{(: where item is T member :) }} \\
	  \textit{(: skip 1st-then-2nd item sequence to generate other possible paths :)} \\
		\eIf{idx = 0}{
			nexIdx = 2			  \\
		} {
			nexIdx = idx+1    \\
		} 				
		\For{nexIdx to length(T)}{
				leafItem $\leftarrow$ item[nexIdx]  \\ 
			\textit{(:if leaf item is descendant of its previous items:) } \\
		 \If{leafItem $\in$ path/* } 
		   {  			
				leafItem/@counter++}
	  }
		\textit{\footnotesize{(: repeat starting with next item of T as parent of the path :)}}\\
		update-other-paths(T, nexIdx)		
	}
}
\caption{Updating other paths\label{alg:step4}}
\end{algorithm}

\subsubsection{Example} 
\label{sec:example}
\begin{figure}[htbp]
\begin{center}
\subfigure[After inserting T1 (A B C D)]{\includegraphics[scale=0.6]{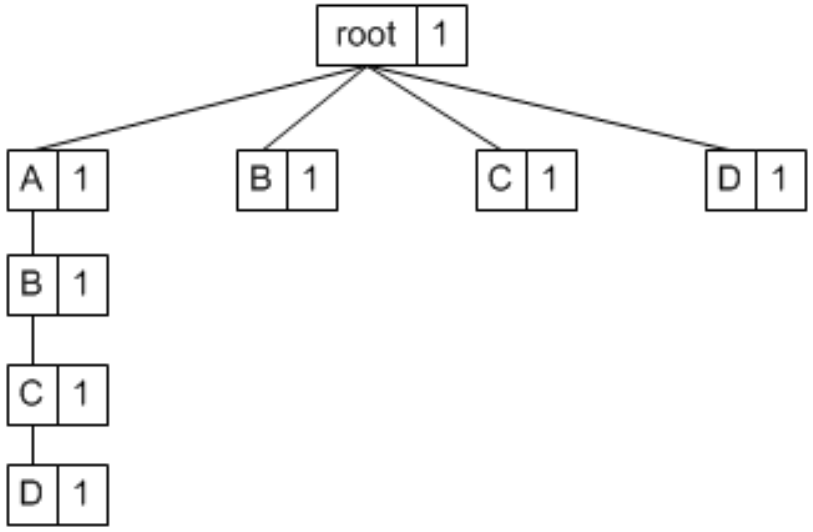}}
\hspace{12pt}%
\subfigure[After inserting T2 (C E)]{\includegraphics[scale=0.6]{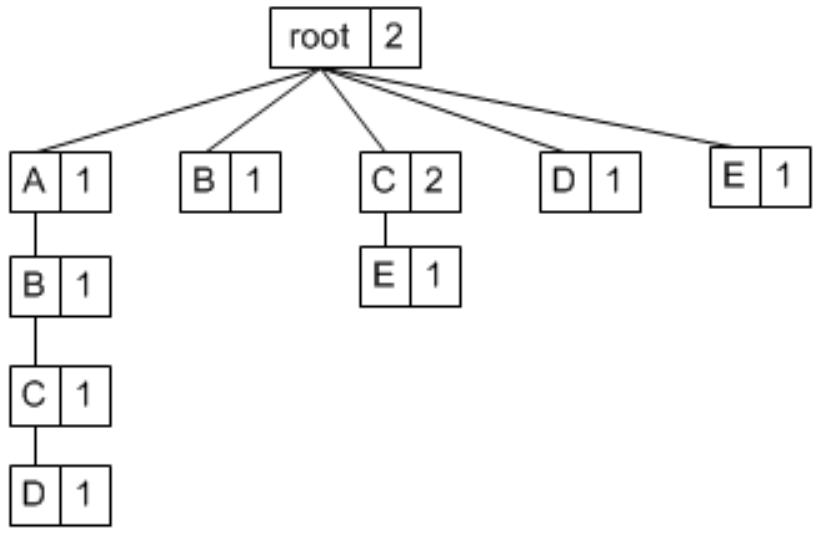}}
\\
\subfigure[After inserting T3 (B C)]{\includegraphics[scale=0.6]{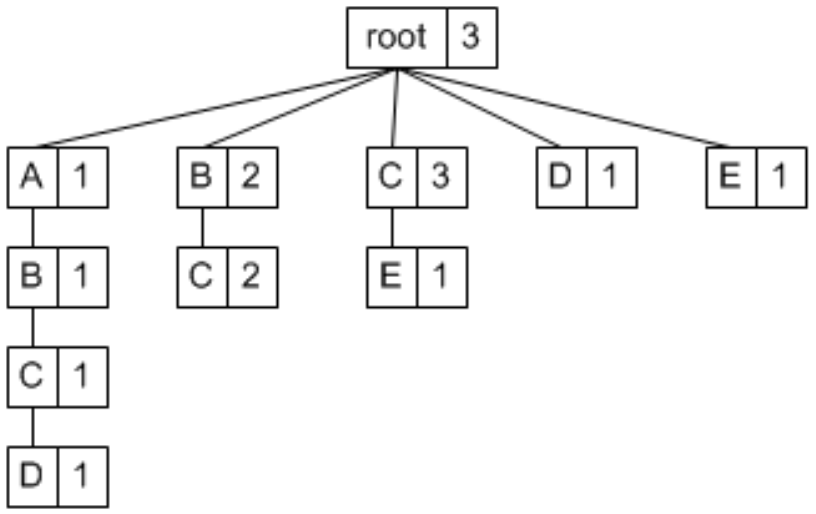}}
\hspace{12pt}%
\subfigure[After inserting T4 (C D E)]{\includegraphics[scale=0.6]{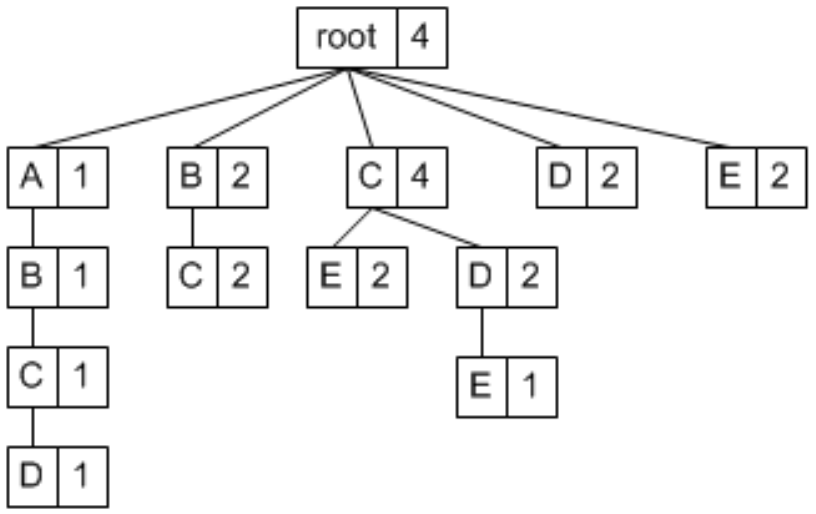}}
\\
\subfigure[After inserting T5 (B C D)]{\includegraphics[scale=0.6]{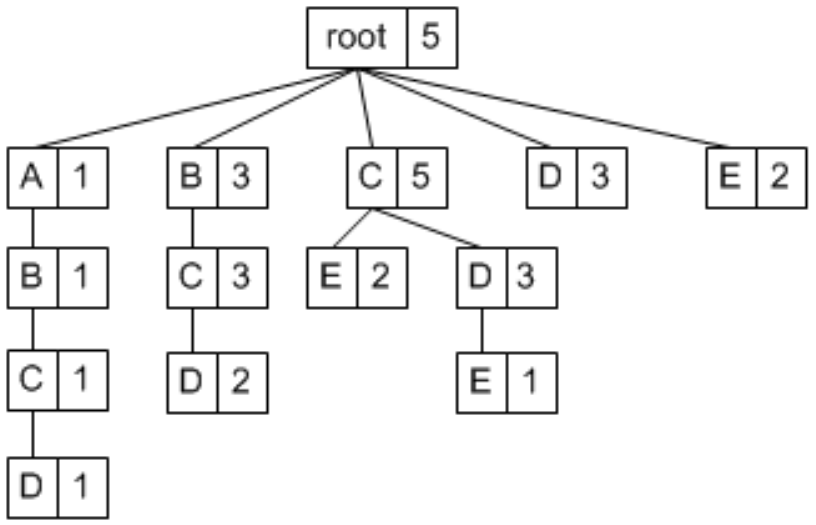}}
\hspace{12pt}%
\subfigure[After inserting T6 (A C E)]{\includegraphics[scale=0.6]{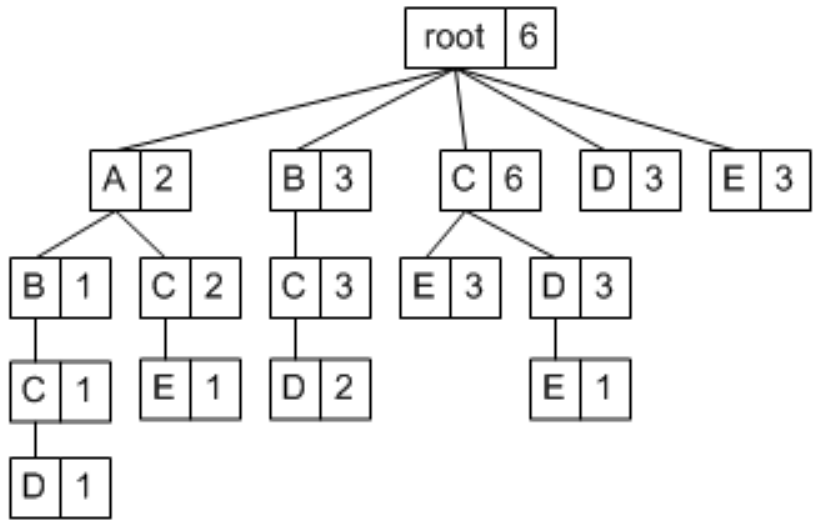}}
\\
\caption{\label{fig:tree} Constructing the FXT document.}
\end{center}
\end{figure}

Figure \ref{fig:tree}(a-f) shows the four steps to construct the FXT by inserting transactions given in section \ref{sect:loggedEvents}. 
Because steps 1 and 2 are always applied directly for all transactions
, we focus on how steps 3 and 4 are applied.  

In figure \ref{fig:tree}(a) and figure \ref{fig:tree}(b), step 3 creates the paths ``root/A/B/C/D" and ``root/C/E", respectively. Step 4 is not evaluated, because there are no other paths available. 
In figure \ref{fig:tree}(c), in order to initialize counter of item ``C" according to step 3, the algorithm detects item ``C" as child of item ``B" in the path ``root/A/B/C/D". Thus, counter of item ``C" in the existing path is incremented to become an initialization value of item ``C" in the new path ``root/B/C".    
In figure \ref{fig:tree}(d), step 3 initializes counter of item ``D" at 2, because item ``D" already exists as child of item ``C" in the path ``root/A/B/C/D". But, step 4 detects other path ``root/C/E" in the FXT, thus the counter of item ``E" is incremented.     
In figure \ref{fig:tree}(e), step 3 initializes the counter of item ``D" at 2, because item ``D" already exists as grandchild or child of items ``B" and ``C", respectively in the path ``root/A/B/C/D". Moreover, step 4 detects portion of other path ``root/C/D" in the FXT, thus counter of item ``D" is incremented. 
In figure \ref{fig:tree}(f), step 3 initializes the counter of item ``C" at 2, because item ``C" already exists as grandchild of item ``A" in the path ``root/A/B/C/D". Also, step 4 detects other path ``root/C/E" in the FXT, thus counter of item ``E" is incremented. 
Finally, the constructed FXT is as follows.

%

{
\myfontsize
\begin{Verbatim}[frame=lines, xleftmargin=5mm]
  <? xml version="1.0" >
  <root counter = "6">
    <A counter = "2">
      <B counter = "1">
        <C counter = "1">
          <D counter = "1"/>
        </C>
      </B>
      <C counter = "2">
       <E counter = "1"/>
     </C>
   </A>
   <B counter = "3">
     <C counter = "3">
       <D counter = "2"/>
     </C>
   </B>
   <C counter = "6">
     <E counter = "3"/>
     <D counter = "3">
       <E counter = "1"/>
     </D>
   </C>
   <D counter = "3"/>
   <E counter = "3"/>
 </root>
\end{Verbatim}	
}

\section{Mining Frequent Itemsets and Association Rules using XQuery}
\label{sect:miningAR}

The main objective of constructing the FXT is to mine frequent itemsets and association rules easily using the XQuery language. Frequent itemsets are queried by traversing the FXT from breadth nodes to specific nodes (portion of paths), or to leaf nodes (complete paths). 
Frequent itemsets are filtered using a statistical measure called \textit{support}. Support measures the proportion of transactions that contains a specific item (or itemset). A frequent itemset is an itemset whose support is greater than some user-specified minimum support. Frequent itemsets satisfy the Apriori property, which states that if a given portion of path does not satisfy minimum support, then neither will any of its descendants \cite{agrawal94:vldb}. 
Examples for retrieving the support of items and itemset from the FXT follow.

\vspace{2pt}
$Support(A) \hspace{25pt} = \frac{root/A/@counter}{root/counter}$  

\vspace{2pt}
$Support (B, C, D)  = \frac{root/B/C/D/@counter}{root/counter}$  

\vspace{2pt}
\textit{\textbf{Example}:} The following example introduces the function for generating frequent itemsets 
from the FXT that constructed in main example (section \ref{sec:example}).


{
\scriptsize
\begin{Verbatim}[frame=lines]
declare variable $input := doc("tree.xml")/root;
declare variable $rootCounter := $input/@counter;

declare function local:getFrequentItemsets($parent as xs:string, 
$element as element(*, xs:untyped), $minSupport as xs:decimal) {

let $path := concat($parent,'/',name($element))
 where $element/@counter div $rootCounter>=$minSupport
return
 (<frequent path="{$path}" count="{$element/@counter}" 
  support="{$element/@counter div $rootCounter}"/>,

 for $child in $element/*
 return
  local:getFrequentItemsets ($path, $child, $minSupport)) };

(: call the function :) 
for $child in $input/*
return
 local:getFrequentItemsets("", $child, 0.25)
\end{Verbatim}
}

The result of calling previous function to get frequent itemsets with min support=0.25 is as follows. 

{
\scriptsize
\begin{Verbatim}[frame=lines, xleftmargin=5mm]
<frequent path="/A" count="3" support="0.375"/>
<frequent path="/A/B" count="2" support="0.25"/>
<frequent path="/A/B/D" count="2" support="0.25"/>
<frequent path="/A/C" count="2" support="0.25"/>
<frequent path="/B" count="4" support="0.5"/>
<frequent path="/B/C" count="3" support="0.375"/>
<frequent path="/B/C/D" count="2" support="0.25"/>
<frequent path="/C" count="6" support="0.75"/>
<frequent path="/C/E" count="3" support="0.375"/>
<frequent path="/C/D" count="3" support="0.375"/>
<frequent path="/D" count="4" support="0.5"/>
<frequent path="/E" count="4" support="0.5"/>
\end{Verbatim}
}

Association rules have been first introduced in the context of retail transaction databases \cite{agrawal94:vldb}. An association rule is an implication of the form \textit{X}$\Rightarrow$\textit{Y}, where the rule \textit{body X} and \textit{head Y} are subsets of the set \textit{I} of \textit{items (I ={$i_1, i_2, . . . , i_n$})} within a set of \textit{transactions D} and $X \cap Y = \Phi$. A rule $X \Rightarrow Y$ states that the transactions \textit{T} that contain the items in \textit{X} are likely to also contain the items in \textit{Y}. Association rules are characterized by two measures: the \textit{support}, which measures the proportion of transactions in \textit{D} that contain both items \textit{X} and \textit{Y}; and the \textit{confidence}, which measures the proportion of transactions in \textit{D} containing items \textit{X} that also contain items \textit{Y}. $Confidence(X \Rightarrow Y)$ can be expressed as the conditional probability $p(Y |X)$. Thus, we define:

{
\vspace{4pt}
\noindent
$Support(X \Rightarrow  Y) \hspace{5pt} = \frac {count(X \cup Y)} {N_{trans}} = \frac {root/X/Y/@counter}{root/counter} \hspace{12pt} (1)$

\vspace{2pt}
\noindent
$Confidence(X \Rightarrow  Y) = \frac{support(X \Rightarrow Y)}{support(X)}$

$ \hspace{70pt}  = \frac{count(X \cup Y )}{count(X)} = \frac{root/X/Y/@counter}{root/X/@counter} \hspace{10pt} (2)$
}

\vspace{2pt}
\textit{\textbf{Example}:} The following example introduces the XQuery function for generating a set of association rules from the FXT constructed in main example (section \ref{sec:example}). 


{
\scriptsize
\begin{Verbatim}[frame=lines]
declare function local:generateRules($parent as xs:string, 
$x as element (*, xs:untyped), $min_sup as xs:decimal, 
$min_conf as xs:decimal) {

let $path_x := concat($parent,'/',name($x)) 
return
 (for $y in $x/*
  let $y_given_x := name($y)
  let $support_xy := $y/@counter div $rootCounter
  let $support_x := $x/@counter div $rootCounter
  let $confidence := $support_xy div $support_x
  where $support_xy >= $min_sup and $confidence >= $min_conf
 return 
  (<rule body="{$path_x}" head="{$y_given_x}" 
	support="{$support_xy}" confidence="{$confidence}"/>,
  local:generateRules($path_x, $y,$min_sup, $min_conf)))  };

(: call the function :) 
for $child in $input/*
return
 local:generateRules("", $child, 0.25, 0.5)
\end{Verbatim}
}

The result of calling previous function to get association rules with min support=0.25 and min confidence=0.5 is as follows. 

{
\scriptsize
\begin{Verbatim}[frame=lines, xleftmargin=5mm]
<rule body="/A" head="B" support="0.25" confidence="0.666"/>
<rule body="/A/B" head="D" support="0.25" confidence="1"/>
<rule body="/A" head="C" support="0.25" confidence="0.666"/>
<rule body="/B" head="C" support="0.375" confidence="0.75"/>
<rule body="/B/C" head="D" support="0.25" confidence="0.666"/>
<rule body="/C" head="E" support="0.375" confidence="0.5"/>
<rule body="/C" head="D" support="0.375" confidence="0.5"/>
\end{Verbatim}
}

Note that, it is possible to apply other XQuery functions to discover some statistics or mine more association rules. For instance, to query the reverse $rule (Y\Rightarrow X)$, a function is firstly required to sort the rule body and the rule head alphabetically in order to calculate the $support (Y\Rightarrow X)$, whereas 

 $count(Y\cup X) = count(X\cup Y) = root/x/y/@counter$. 
But, when calculating the $confidence (Y\Rightarrow X)$, the rule body does not change in the denominator, i.e., $support(Y)$, see equation (2).
	
\section{Performance Study}
\label{sect:performance}

We have implemented the FXT construction algorithm using some Java libraries for manipulating XML data structure (i.e., JDom, SAXPath, and Jaxen). Mining frequent itemsets and association rules are performed using the XQuery language. We experimented with different synthetic datasets, starting from 10 transactions to 100K of transactions. The average lengths of transactions are 15 items per transaction.  All experiments are performed on a 2.80 GHz PC with 3 GB RAM, running on Windows 7, with minimum Java heap size 128 MB and maximum Java heap size 512 MB.    

\begin{figure*}[htbp]
\begin{center}
\subfigure[]{\includegraphics[scale=0.375]{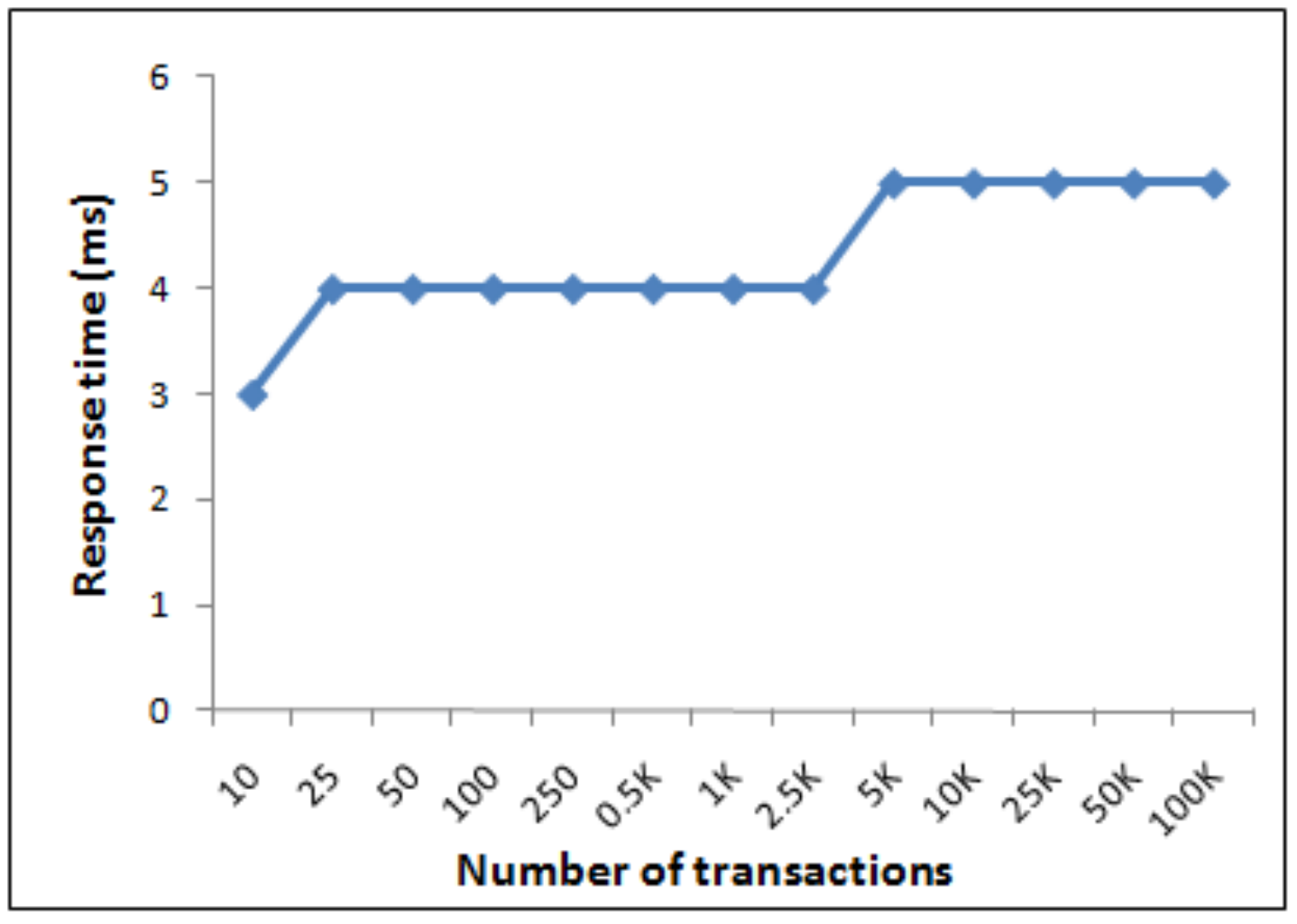}}
\subfigure[]{\includegraphics[scale=0.375]{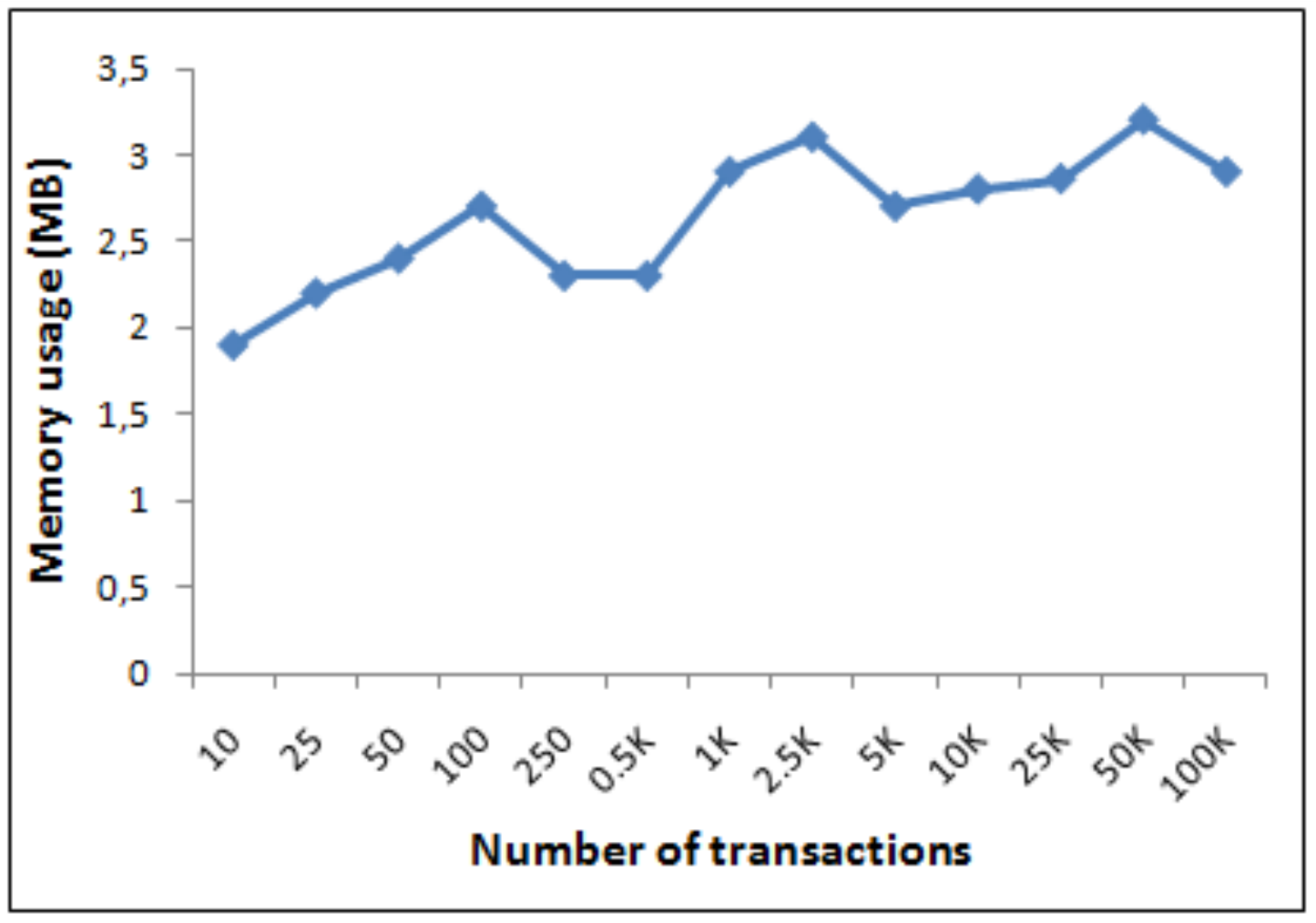}}
\subfigure[]{\includegraphics[scale=0.375]{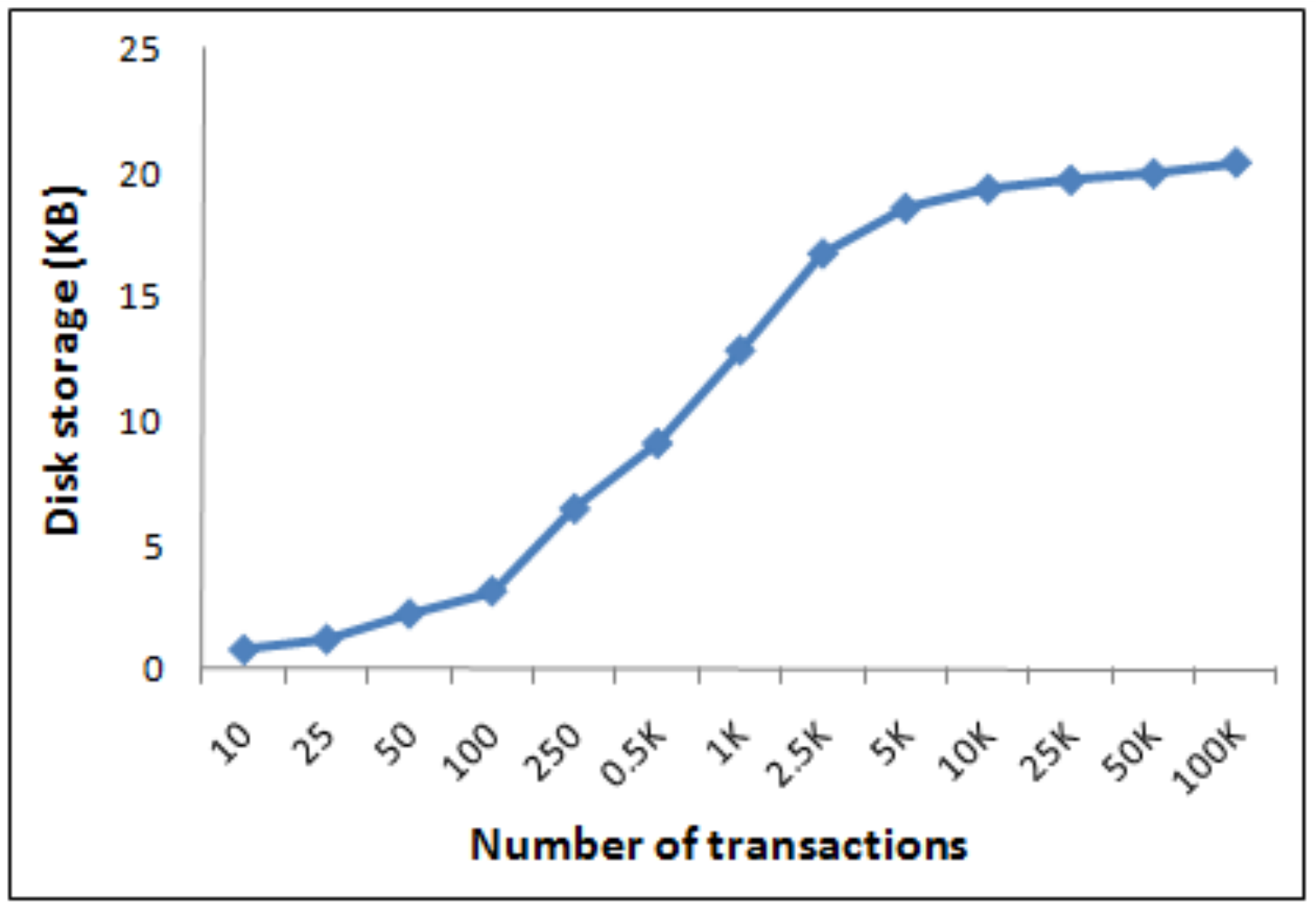}}
\caption{\label{fig:resources} Machine resources for inserting new transaction to FXT}
\end{center}
\end{figure*}

We study the impact of constructing the FXT on the machine resources. Figure \ref{fig:resources}(a) plots CPU time for new transaction insertion given different FXT sizes. It can be easily observed that the CPU runs fast for inserting new transaction even though FXT has large size (e.g., it takes 5ms to insert new transaction into a 100K FXT size). Likewise, figure \ref{fig:resources}(b) plots memory usage, 
it can be observed that our algorithm consumes a small size of memory for new transaction insertion with different FXT sizes. 
Figure \ref{fig:resources}(c) plots disk storage of the FXT document against different sizes of transactions. As shown in the figure, although the increasing relationship, the required storage remains small. Due to the FXT compact structure, the repeated or similar insertions of transaction need to only update item counters without consuming further storage space. 

\begin{figure*}[htbp]
\begin{center}
\subfigure[]{\includegraphics[scale=0.55]{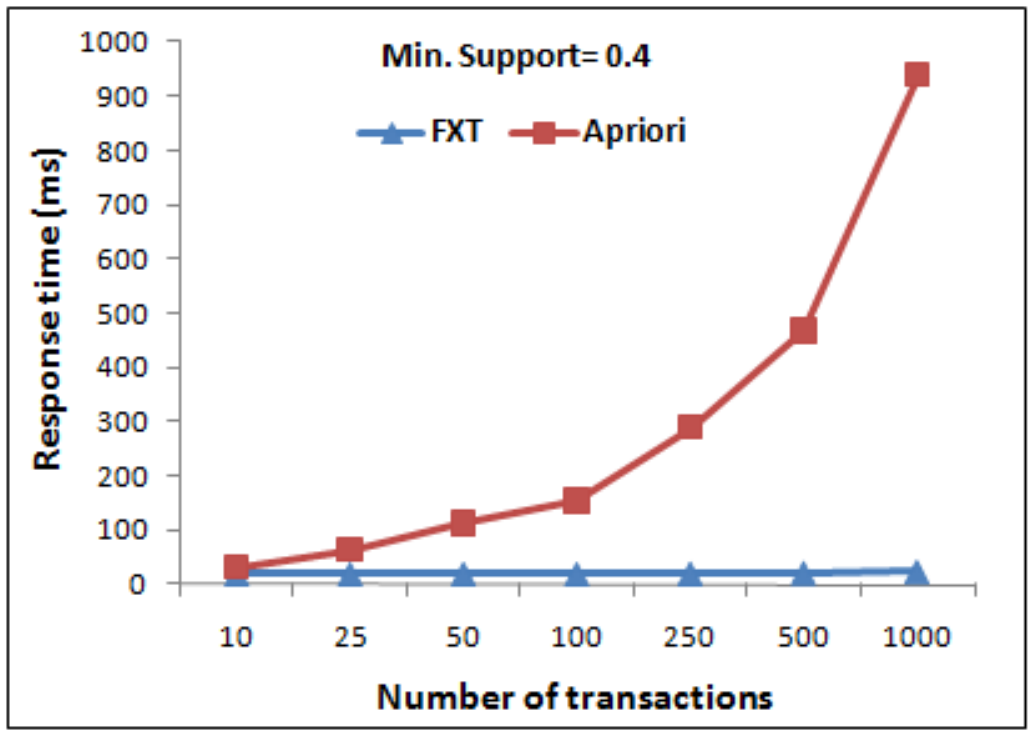}}
\subfigure[]{\includegraphics[scale=0.55]{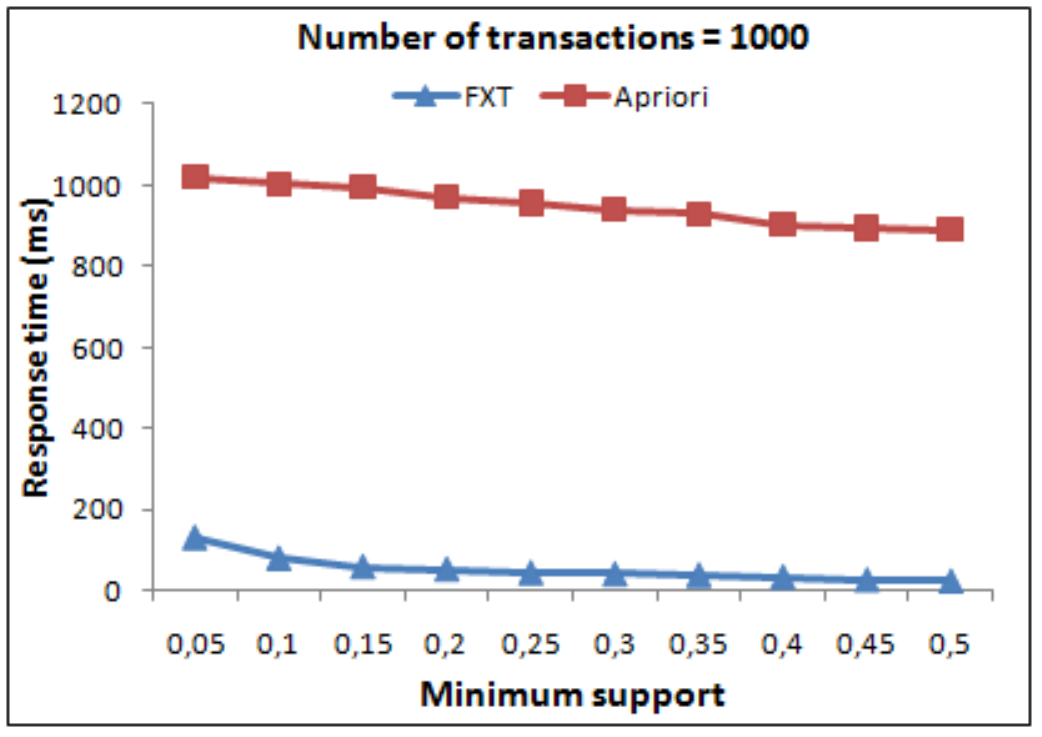}}
\caption{\label{fig:fxt_apriori} Performance comparison between FXT and Apriori algorithm}
\end{center}
\end{figure*}

Since we are interested in mining XML data using XML technologies, to the best of our knowledge there is only one most related work (i.e., implementation of Apriori algorithm using XQuery \cite{wan04:dmwi}). The Apriori algorithm always deals with static database of transactions. Figure \ref{fig:fxt_apriori}  shows the performance comparison between our algorithm and the XQuery-based implementation of Apriori, for mining association rules from XML using XQuery. It shows that our algorithm is always providing better performance than Apriori, specifically for larger amount of transactions (see figure \ref{fig:fxt_apriori}(a)), and also for different values of minimum support (see figure \ref{fig:fxt_apriori}(b)). Apriori generates frequent itemsets and association rules each time from scratch, while our algorithm construct the FXT incrementally. Then frequent itemsets and association rules can be queried directly at any time from the FXT.
Moreover, FXT is very compressed if compared with transactions document of Apriori algorithm.  

Finally, we conclude that our algorithm is very efficient to consume resources. It can also mine frequent itemsets and association rules against different support and confidence values, without reconstructing its FXT from scratch that results in a better performance. Additionally, FXT performance is better than XQuery-based Apriori implementation.  
\section{Conclusions}
\label{sect:conclusion}

In this paper, we propose an incremental approach for mining association rules from XML logged events. Our approach applies an incrementing breath-then-depth algorithm, for constructing a novel frequency XML-based tree structure. The algorithm composes of four steps for inserting transaction into the tree. The constructed tree can be directly queried using XQuery language for retrieving frequent itemsets and association rules, without applying complex data mining techniques. Our algorithm handles incremental logged events. Thus, it is featured with a single-pass of dataset, incremental processing of transaction, compressed structure of the tree, fast for inserting new transactions, fast for querying frequent itemsets or association rules, and efficient to limited resources. These features are validated by implementing the algorithm and experimenting its performance. 

In future, we aim at mining association rules from logged events taking into account their real-time of logging, and discovering the relationships among events against their logged real-time. Moreover, we intend to apply our algorithm for mining XML events that logged from our data integration platform \cite{salem10:icwmi}. This algorithm can be used to discover interesting knowledge, in order to maintain, automate, and re-activate the workflow behavior of the ETL tasks.

\bibliographystyle{abbrv}
\bibliography{mybibfile}

\end{document}